# Hosting Second Order Exceptional Point in an All-lossy Dual-Core Photonic Crystal Fiber

**Shamba Ghosh**[1, 2], **Arpan Roy**[3], **Bishnu P. Pal**[4] **and Somnath Ghosh**[4, *]

[1] *Department of Physics, Indian Institute of Technology Jodhpur, Rajasthan 342030, India*
[2] *Department of Metallurgical and Materials Engineering, Indian Institute of Technology Jodhpur, Rajasthan 342030, India*
[3] *Institute of Radio Physics and Electronics, University of Calcutta, Kolkata-700009, India*
[4] *Department of Physics, École Centrale School of Engineering, Mahindra University, Hyderabad, Telangana, 500043, India*
*\*Corresponding author: somiit@rediffmail.com*

**Abstract:** We report an all-lossy index-guided dual-core photonic crystal fiber (PCF) that hosts a second-order exceptional point (EP) in the system's parameter space. By appropriately selecting a parametric encirclement scheme around the EP, the interaction between the coupled modes has been studied, and the mode conversion is subsequently observed. © 2025 The Author(s).

## 1. Introduction

An EP is a unique topological singularity that appears in the parameter space of non-Hermitian systems, where both the eigenvalues and eigenstates of the system's Hamiltonian coalesce simultaneously [1-5]. The interplay between non-Hermitian components, such as gain-loss, and the topological properties of an open system governs the interaction among its complex energy levels, leading to an avoided resonance crossing-type phenomenon [1]. A gradual variation of the non-Hermitian parameters around an EP results in an adiabatic transition of the eigenvalues. Recently, there has been growing interest in open photonic systems that host EPs, given their potential for diverse applications, including optical isolators [2], asymmetric mode switching [3], and ultra-sensitive sensors [4]. While gain-loss-assisted PCF has been explored for hosting EP [5], their fabrication remains a practical challenge. Fortuitously, an all-lossy PCF offers a more feasible alternative. In this work, we introduce an all-lossy dual-core PCF segment, departing from the conventional gain-loss framework used in our prior study [5]. The decision to adopt an all-lossy system is motivated by practical considerations because incorporating gain requires additional components, precise doping with active materials (e.g., Er, Yb), and optical pumping, which makes the process costly and vulnerable to instabilities and fluctuations. In contrast, introducing loss simply involves doping with a lossy material, where the magnitude of loss can be precisely controlled through appropriate tailoring of the dopant concentration [4]. This provides a cost-effective and low-noise alternative while retaining the intrinsic advantages of fiber-based platforms. Here, we propose a dual-core PCF that supports two quasi-guided modes at a wavelength of $1.55\ \mu m$. By implementing a customized, and disproportionate loss distribution in the two cores, we investigate the mode-mode interaction and host an EP in a 2D parameter space. Hosting of an EP in this all-lossy microstructure fiber geometry presents a promising avenue for highly sensitive EP-based sensing and lays the foundation for the development of next-generation photonic devices.

## 2. Results and discussions

We considered a dual-core index-guided PCF segment with a cladding refractive index ($n_c$) of 1.45. This fiber consists of air holes, which are arranged in a hexagonal pattern in the cladding and run through the length of the fiber. As shown in **Fig. 1(a)**, the structure has two defects present in the hexagonal pattern, which are separated by a central air hole. These two defects are the two guiding cores of the fiber as light propagation is confined within these cores, enabling guided optical modes. The pitch and the air hole radius are optimized to ensure that the structure supports only two guided modes, namely $\psi_1\ and\ \psi_2$, as depicted in **Fig. 1(b)**. In this passive system, non-Hermiticity is incorporated by implementing customized loss profiles in both the cores and cladding, thereby enabling control of the interaction between the two modes. A fixed amount of loss $\gamma_{FL}(= 5 \times 10^{-4})$ is incorporated in the cladding, and the magnitude of loss distribution in two separate cores is modulated by two independent parameters: loss coefficient ($\gamma$) and fractional-loss ratio ($\tau$). The resulting complex refractive index distribution is defined as follows: left core: $n_L = n_c + i(\gamma_{FL} - \gamma)$, right core: $n_R = n_c + i(\gamma_{FL} + \gamma/\tau)$ and cladding: $n_c + i\gamma_{FL}$, as depicted in **Fig. 1(c)**. This configuration enables precise control of the non-Hermitian interactions between the modes within the PCF.

Mutual coupling occurs between $\psi_1$ and $\psi_2$, as the system transits into a non-Hermitian regime. When the tuning parameters $(\gamma, \tau)$ are varied the left core experiences less loss whereas the right core experiences higher loss, ensuring that the system remains gain-free. The loss parameter $\gamma$ is systematically varied within the range of 0 to $5 \times 10^{-4}$, for several $\tau$ values, and the $n_{eff}$-values of the modes are calculated using Optiwave® simulation software. These values represent their effective modal indices and are equivalent to the eigenvalues of the Hamiltonian. By examining the paths of complex effective index of the modes, we investigate the interactions between two coupled modes. The variations of the real and imaginary parts of $n_{eff}$- values of the two modes are plotted against increasing $\gamma$ for a fixed $\tau = 1$, with $\psi_1$ represented by the blue curve and $\psi_2$ by the red curve, as shown in **Fig. 1(d)**. The real part of $n_{eff}$, indicated by solid lines, coalesces at $\gamma = 0.00018$, while the imaginary part, represented by dotted lines, bifurcates at the same point. This simultaneous coalescence of real parts and bifurcation of imaginary parts is a key signature of an EP. Consequently, it validates that an EP is hosted in the 2D parameters space between $\psi_1$ and $\psi_2$ at (0.00018,1).

To confirm the second-order branch point behavior of the EP, it is encircled within the $(\gamma, \tau)$- plane using a closed elliptical trajectory. As shown in **Fig. 1(e)**, this loop is defined by the parametric equations, $\gamma(\phi) = \gamma_0 \sin(\phi/2)$ and $\tau(\phi) = \tau_{EP} + r \sin\phi$, where the characteristic parameter $r = 0.5$, the adjustable angle $\phi$ varies within the range of 0 to $2\pi$ and $\gamma_0 = 0.0005 > \gamma_{EP}$. The direction of variation is determined by the choice of $\phi$, where a clockwise (CW) encirclement is defined by $\phi$ varying from 0 to $2\pi$, while an anticlockwise (ACW) encirclement follows a $2\pi$ to 0 variation. The complex $n_{eff}$ values of the modes are calculated as the loss parameter varies along the loop in the CW direction, and their trajectories are plotted in the complex $n_{eff}$ plane, as shown in **Fig 1(f)**. It is observed that the $n_{eff}$ associated with $\psi_1$ and $\psi_2$ interchange their original positions during the encirclement of the EP. This flip-of-state phenomenon between the modes verifies that the EP inherently exhibits branch point behavior.

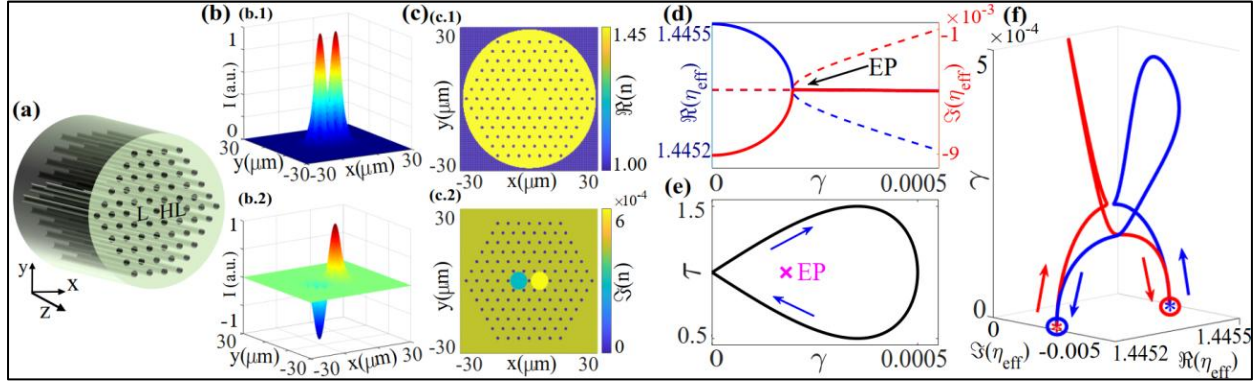

**Fig. 1 (a)** Schematic of the proposed all-lossy dual-core index-guided PCF segment with the transverse cross-section in xy-plane and the propagation direction along the z-axis (L → Loss, HL → High loss). **(b)** Normalized intensity profile of two supported guided modes $\psi_1$(b.1) and $\psi_2$(b.2). **(c)** The refractive index profile in the transverse $xy$ plane, where the real part (c.1) is fixed and the imaginary part (c.2) is shown for a specific set of parameters $\gamma = 1.8 \times 10^{-4}$, $\tau = 1$, and a fixed loss $\gamma_{FL}(= 5 \times 10^{-4})$ is added to the entire region. **(d)** Trajectories of the complex $n_{eff}$ values of two quasi-guided modes varying with the loss parameters $\gamma$. For $\tau = 1$, the variation of the $n_{eff}$ of $\psi_1$ (blue curve) and $\psi_2$ (red curve) shows that the real parts (solid line) coalesce while the imaginary parts (dotted line) bifurcate at $\gamma = 0.00018$, simultaneously. **(e)** The chosen parametric loop in the $(\gamma, \tau)$ plane to encircle the EP (pink cross). **(f)** Variations of $n_{eff}$-values for the modes $\psi_1$ (blue curve) and $\psi_2$ (red curve) during the parametric encirclement illustrates the switching between the modes due to encircling the EP. The arrows of respective color indicate the directions of evolution, stars mark the starting points, and circles denote the ending points.

## 3. Summary

In summary, we present the first observation of a second-order exceptional point in an all-lossy index-guided dual-core PCF, which is achieved by designing an optimized fiber geometry that supports only two guided modes and incorporating an unbalanced loss in two separate cores. By examining an appropriately designed parametric loop around the EP, we investigate the inherent branch point property of EP. For the selected encirclement processes, the mutual switching between the $n_{eff}$-values of the coupling modes is observed. Additionally, the resulting modal propagation dynamics, which can be realized through the dynamic encirclement of the EP, will be explored in further detail. This all-lossy system eliminates instabilities, fluctuations, and high quantum noise associated with gain media and simultaneously circumvents the complexity and cost to achieve the controlled gain profile distribution. Our proposed all-lossy PCF should be relatively easy to fabricate using modern fabrication techniques, making it an ideal platform for applications in optical communication devices, mode conversion, and quantum sensing.

**Funding:** Shamba Ghosh acknowledges the financial support from the Ministry of Education, Government of India.